\documentstyle[prl,aps]{revtex}
\begin{document}
\draft
\title{
Direct current generation due to wave mixing in semiconductors}
\author{
Kirill N. Alekseev\cite{email1}, Mikhael V. Erementchouk}
\address{
Kirensky Institute of Physics, Russian Academy of Sciences, \\
Krasnoyarsk 660036, Russia}
\author{Feodor V. Kusmartsev\cite{email2}\thanks{also Landau 
Institute for Theoretical Physics,
Russian Academy of Sciences, Moscow 142432, Russia}
}
\address{
School of Mathematical and Physical Sciences , \\
Loughborough University, Loughborough LE11 3TU, UK}
%\date{}
\maketitle
\begin{abstract}
We describe a novel effect of the generation of direct current which may
arise in semiconductors or semiconductor microstructures
due to a mixing of coherent electromagnetic radiations of commensurate 
frequencies. The effect is, in essence, due to a nonparabolicity of 
the electron energy bands and is stronger in systems where this 
nonparabolicity is greater. We have made 
exact calculations in the framework of the Kane model, applicable 
to narrow gap semiconductors and the tight-binding model which we 
employ for a description of a semiconductor superlattice.
\end{abstract}

The problem of emission and reception of electromagnetic radiation has 
attracted the attention of the scientific community for a long time. 
Good sources
of coherent electromagnetic radiation, its receivers and detectors
exist for the radio-frequency, microwave and optical ranges of the spectrum.
Many of these devices are based on semiconductor technology. Nowadays,
the terahertz range ($0.1$ - $1$ THz, submillimetre wavelengths)
is practically the last unexploited {\it terra incognita}.
\par
In 1970, Esaki and Tsu \cite{Esaki-ibm}
made a pioneering suggestion  to use
the semiconductor superlattice (SSL) 
for generation of Bloch oscillations of frequency
$\omega_B=e a E_0/\hbar$, where $E_0$ is a constant electric field
applied along the axis of a SSL with a spatial period $a$. For typical SSLs
and bias $E_0$ of $1$ - $10$  kV/cm, the Bloch
frequency belongs to the THz range \cite{Esaki-ibm}. This can be
used for the generation of THz radiation \cite{THz-emission}.
The work of Esaki and Tsu \cite{Esaki-ibm} stimulated enormous
theoretical activity devoted to the interaction of a high-frequency
electric field with a SSL \cite{Bass1,Bass2}. Moreover, recent 
progress 
in the development of THz radiation sources and coupling
techniques allow the systematic experimental studies of many of the
associated nonlinear effects \cite{SSL-exp}. Amongst others, one of the most
interesting
suggestions was that of Esaki and Tsu \cite{Esaki-apl} and Romanov
\cite{Romanov-opt} to use the SSL as a new, artificial, nonlinear material
for electromagnetic wave mixing and harmonic generation. The theory
of wave mixing in SSLs, based on a solution of the Boltzmann equation, 
has been developed by Romanov and co-workers \cite{Romanov-ftt,Bass1}.
\par
Very recently, in cooperation with the Urbana group, we showed
\cite{Alekseev-prl} that the effect inverse to the Bloch 
oscillations in SSL is
possible, namely, an alternating field without constant bias can create
Bloch oscillations in a single miniband SSL. The appearance of 
induced Bloch oscillations means a spontaneous creation of 
constant voltage and corresponding direct current (DC) 
\cite{Alekseev-prl}, that can be considered as a rectification of 
the THz field in SSLs. The rectification
in \cite{Alekseev-prl} results from an interplay of dissipation
(scattering
of ballistic electrons with impurities and phonons) and generation of
a self-consistent electric field along the SSL axis, which is a strongly
nonlinear effect. Moreover, the generation of DC in \cite{Alekseev-prl}
is
a counterpart of chaos, which arises for the same {\it ac} field strength and
frequency but for much weaker dissipation \cite{Alekseev-prb}.
\par
Independently, Goychuk and H\"{a}nggi \cite{Goychuk-epl} 
suggested another scheme of
quantum rectification using a wave mixing of an alternating electric
field
and its second harmonic in a single miniband SSL. The approach of
\cite{Goychuk-epl} is based on the theory of quantum
ratchets (see \cite{Goychuk-prl} and references therein) and
therefore the necessary conditions for the appearance of DC
include a dissipation (quantum noise) and an extended periodic
system. 
\par
In this work we find an even more general
effect of DC generation, or, a rectification
at wave mixing, which may arise in any semiconductor
or semiconductor microstructure. The origin of this effect
is related to neither dissipation nor quantum noise but
rather to a nonparabolicity of the energy spectrum. This 
nonparabolicity is always present in any semiconductor or semiconductor 
structure \cite{Bir,Bass2}.
The values of the nonparabolical energy terms
may take their largest values in semiconductors 
with wide bands or small effective masses. 
It seems that the size of the gaps is not important,
however, semiconducors with wide gaps only have small concentrations of 
current carriers unless, of course, they are doped. 
Hence, the best candidates for the described effect are semiconductors
having wide bands and narrow gaps or doped semiconductors 
such as InSb and other analogous compounds.
Indeed, wave-mixing has already been observed in such narrow-gap 
semiconductors \cite{Patel,Wynne,Gorky-exp}; 
its mechanism being mainly related to the 
nonparabolicity of the energy band
\cite{Patel,Wynne,Gorky-exp,Wolff,Gorky-theor}.
\par
We also consider another system, SSLs,
which is now under intensive  experimental investigation
and  the  possibility of wave mixing there has also been
suggested. 
The description
of a SSL in the framework of the tight binding model allows
exact solutions to be obtained.  Moreover, since 
the  nonparabolical energy term in a SSL takes
smaller values than for normal semiconductors, the estimation
of the DC generation effect for this system gives a lower bound to
the value of the DC which may be observed in semiconductors. 
We show that the main features of the novel effect, the 
generation of  DC due to the mixing of harmonics, 
may be well described and understood within
the standard semiclassical Boltzmann equation approach
\cite{Romanov-ftt,Bass1}.
We demonstrate that the value of DC is strongly dependent on the
product of a characteristic scattering time, $\tau$, and field
frequency, $\Omega$. In the experimentally relevant case of weak field
strength, the expression for DC can be represented in the form of
the generalized Ohm's law with prefactor dependent on $\Omega\tau$.
The value of DC is maximal at $\Omega \simeq \tau^{-1}$ (namely,
$\Omega\tau\approx 0.7$) and it decreases quadratically for both small
and large values of $\Omega\tau$.
\par
The novel effect we find is consistent with results
obtained  previously in a quantum approach \cite{Goychuk-epl}
for a special case of quantum noise.
However, our predicted effect is much more general and has a much 
simpler physical picture associated with wave mixing 
in nonlinear media arising due to a nonparabolicity
of the electron bands in semiconductors.
In this new framework the results of
\cite{Alekseev-prl,Goychuk-epl}, which at first glance
look very different, may be unified. The generality of the novel
effect  highlights new directions for various different applications 
and uses of semiconductor technology.
\par
Now, we demonstrate how the  rectification effect
arises due to a nonparabolicity of the spectrum of the semiconductors.
Note that, although the dissipative property of a semiconductor or 
microstructure
has an influence on the effect, it only plays a secondary role.
To study the generation of a finite DC let us 
consider, for simplicity, a cubic
semiconductor subjected to the electric mixing harmonic fields
%Eq(1)--------------------------
\begin{equation}
\label{E_hm}
E(t)=E_1\cos(\Omega t)+E_2\cos(2\Omega t+\phi),
\end{equation}
%------------------------------------
With the aid of an effective mass method \cite{Bir}
the energy-momentum dispersion relation of a nondegenerate
cubic semiconductor in the vicinity of the bottom of the conduction band
may be represented as (see, for comparison, Ref. \cite{Bass2})
%Eq(2)--------------------------
\begin{equation}
\label{effective-mass}
\varepsilon(p_x,p_y,p_z)=\frac{p^2 }{2 m}+\frac{\eta}{4}
(p_x^4+p_y^4+p_z^4),\quad
\frac{1}{m}=\left.\frac{\partial^2\varepsilon}{\partial p_i^2}
\right|_{p_i=0}, \quad
\eta=\left.\frac{1}{6}\frac{\partial^4\varepsilon}{\partial p_i^4}
\right|_{p_i=0},
\end{equation}
%------------------------------------
where $p^2=p_x^2+p_y^2+p_z^2$, $m$ is an effective mass 
at the bottom of the conduction band, $\eta$ is a
parameter of nonparobolicity. Following \cite{Bass2}, for simplicity,
we consider electron motion only along one direction,
for example, along the ${\bf x}$-direction and therefore, we may
limit ourself to the one dimensional approximation. From now on the index $x$
is suppressed. The electron's velocity is given by 
$v=\partial\varepsilon/\partial p$, the DC related to nonparabolicity is
$j_{\rm dc}=e n\eta\langle p^3\rangle$,
where $n$ is the number of electrons per unit volume and
the angled brackets $\langle\ldots\rangle$ indicate the time averaging over a
period of the electric field $2\pi/\Omega$. For pure ballistic
electron 
motion without scattering by impurities or phonons, the electron
dynamics is determined by the accelerating theorem $\dot{p}=e E(t)$.
Combining this formula with Eqs. (\ref{E_hm}) and
(\ref{effective-mass}), we have
%Eq(3)------------------------------
\begin{equation}
\label{dc-ham}
j_{\rm dc}\propto -e n\eta\frac{3}{8}
\frac{e^3}{\Omega^3} E_1^2 E_2.
\end{equation}
%--------------------------------------------
Next, we will consider two specific examples
where the value of the DC can be expressed
explicitly through the well-established parameters
of the semiconductor energy bands. 
The energy-momentum dispersion relations, $\varepsilon({\bf p})$,
of a cubic semiconductor is usually
obtained within the effective mass method \cite{Bir}. 
For a SSL, the dependence
$\varepsilon({\bf p})$ was obtained within a
tight-binding model \cite{Bass1}.
In both cases, in the weak field limit when the excitation
energy of electrons is small in comparison with
the bandwidth, we may take into account
only the first terms in the energy momentum relation
(see, the dependence (\ref{effective-mass})).
\par
We start with the more transparent case of a SSL.
Here the interplay of dissipation effects and nonparabolicity of 
the miniband may be described exactly. 
Consider the motion of an electron within a single
miniband of the SSL with spatial period $a$ and miniband width $\Delta$.
In the standard tight-binding approximation, we have
%Eq(4)--------------------------
\begin{equation}
\label{tight-binding}
\varepsilon=\frac{\Delta}{2}\left[ 1-\cos\left( \frac{p a}{\hbar}\right)
\right],
\end{equation}
%------------------------------------
where $p$ is the momentum of the electron along the SSL axis. For a 
weak electric field is the electrons oscillate in $p-$space near
the center of the Brillouin zone $|p|\ll\pi\hbar/a$ or at
the bottom of the miniband. Then, from Eq. (\ref{tight-binding}),
we have (\ref{effective-mass}) with $\eta=-a^2/6\hbar^2 m$ and
$m=(2\hbar^2)/(\Delta a^2)$. The DC [Eq. (\ref{dc-ham})] takes the form
$j_{\rm dc} \simeq \frac{e n\hbar}{a m}\xi_1^2\xi_2$,
where $\xi_l=\frac{e a E_l}{l\hbar\Omega}$, ($l=1,2$).
It is easy to see
that the condition of weak nonparabolicity, $|p|\ll\pi\hbar/a$, 
corresponds to $\xi_l\ll 1$.
We turn to the detailed calculations of the DC  taking into account
collisions  of electrons with impurities and phonons,
{\it i.e.} dissipation effects. 
\par
The electron transport properties in narrow miniband SSLs at
temperatures above $40$K are known \cite{Grahn-prb} to be
well described by the semiclassical Boltzmann equation with a constant
relaxation time, $\tau$. 
Starting from a formal
solution of the Boltzmann equation with constant relaxation time,
Romanov and co-workers found the exact expression for a 
time-dependent current $j(t)$ in a tight-binding lattice 
[Eq. (\ref{tight-binding})]
subjected to an electric field with two
frequencies $\omega_1$ and $\omega_2$ \cite{Romanov-ftt,formula}
(see also Appendix \ref{appendix1}).
Taking  $\omega_1=\Omega$, $\omega_2=2\Omega$ and averaging $j(t)$
over the period of the {\it ac} field, $2\pi/\Omega$, we get for the rectified DC
$j_{\rm dc}=\langle j(t)\rangle$ for $t\gg\tau$ the following formula
%Eq(5)--------------------------
\begin{equation}
\label{j_dc-Rom}
j_{\rm dc}=j_0 \sum_{\mu_1,\mu_2=-\infty}^{+\infty}
\sum_{\nu=-\infty}^{+\infty}
\frac{(\mu_1+2\mu_2) x\cos(\nu\phi)+\sin(\nu\phi)}{1+(\mu_1+2\mu_2)^2
x^2}
J_{\mu_1}(\xi_1) J_{\mu_2}(\xi_2) J_{\mu_1-2\nu}(\xi_1)
J_{\mu_2+\nu}(\xi_2),
\end{equation}
%------------------------------------
where $x=\Omega\tau$, $j_0=\frac{\hbar \sigma}{e \tau a}$ with
$\sigma$
being a static, ohmic conductivity along the SSL axis,
$J_{\mu}(\xi)$ is the Bessel function.
\par
The calculation of the DC with the use of Eq. (\ref{j_dc-Rom})
shows that the DC strongly depends on the product $x=\Omega\tau$.
Fig. 1 illustrates the dependence of the DC on $\xi_1$ and $\xi_2$
for different values of $x$. For small $x$, the absolute
value of the DC increases monotonically with an increase of both 
$\xi_1$ and $\xi_2$ (Fig. 1a). The increasing slope of the DC graph 
changes dramatically when the value of $x$ increases.
For instance, this slope can increase by almost
five orders of magnitude with an increase in $x$ of only one order 
(compare
Figs. 1a and 1b). However, for further change of $x$ from
$x=0.2$ (Fig. 1b) to $x=1$ (Fig. 1c), the corresponding increase in 
DC slowsdown and becomes nonmonotonical. The DC reaches its
maximal value for some optimal value of the relationship 
between $\xi_1$ and $\xi_2$. So for the case
$x=1$, the DC is maximal when $\xi_1/\xi_2\simeq 1.0$ 
(see Fig. 1c).
\par
When the electric field amplitudes are small, $\xi_{1,2}\ll 1$, we can use the
Bessel function approximation  $J_n(\xi)\approx (\xi/2)^n (1/n!)$ and 
obtain from (\ref{j_dc-Rom}) the following analytic expression for the
DC (see Appendix \ref{appendix2})
%Eq(6)--------------------------------------------------
\begin{equation}
\label{result1}
j_{\rm dc}=-1.5 j_0  \frac{x^3}{4x^4+5x^2+1} \xi_1^2 \xi_2
\cos\phi
+{\rm O}(\xi^5).
\end{equation}
%-------------------------------------------------------
Substituting usual Drude conductivity $\sigma_{\rm drude}=n e^2 \tau/m$ 
into $j_0$, we get factor $(e n\hbar)/(a m)$, which 
is similar to the prefactor for $j_{\rm dc}$ obtained within colisionless
approximation.
The equation (\ref{result1}) agrees well with the asymptotic
dependence of the DC on the field amplitudes,  $j_{\rm dc}\propto
\xi_1^2 \xi_2 \cos\phi$ as obtained in reference \cite{Goychuk-epl} in
the framework of a different approach. However,
our result (\ref{result1}) also gives the dependence of DC on the
parameter
$x=\Omega\tau$ as well as indicate that the next significant contributions 
to the current appear only when the parameter, $\xi$, is of the order of $\xi^5$.
\par
Let us estimate the value of the predicted DC for the experimental 
conditions of THz field driven SSLs \cite{SSL-exp}. 
For the miniband width
$\Delta\simeq 10$ meV, superlattice period $a\simeq 10$ nm, 
electron density $n\simeq 10^{16}$ ${\rm {cm}^{-3}}$, sample area
$S\simeq 10$ ${\rm {(\mu m)}^{2}}$, electric field of amplitude
$E\simeq 1$ ${\rm kV/cm}$ and frequency $\Omega$ of several THz
($\xi_1=\xi_2\simeq 0.1$) and the characteristic
relaxation time of several picoseconds ($x=\Omega\tau\simeq 1$), we get
the DC $I_{\rm dc}=j_{\rm dc} S\simeq 0.1$ ${\rm \mu A}$.
This value is in very good agreement with an estimate for the DC obtained
by
Goychuk
and H\"{a}nggi \cite{Goychuk-epl} from numerical calculations of the
integral
representations for DC within a different approach based on the 
ratchet theory.
\par
Note that DC density (\ref{result1}) can be represented in the form of the
generalized nonlinear Ohm's law incorporating only total electric field 
$E(t)$ [Eq.(\ref{E_hm})].
Really, using the property $\langle E^3\rangle=3/4 (E_1^2 E_2)\cos\phi$,
we have from (\ref{result1}) the following expression for the DC
%Eq(7)--------------------------------------------------
\begin{equation}
\label{result2}
j_{\rm dc}=-\sigma f(x) \langle\overline{E}^2 E\rangle,\quad
f(x)=\frac{x^2}{4x^4+5x^2+1},\quad \overline{E}\equiv\frac{e a E(t)}
{\hbar\Omega}.
\end{equation}
%-------------------------------------------------------
The nontrivial prefactor $f(x)$ increases as $x^2$ in the limit $x\ll
1$,
reaches its maximal value $f(x^*)\approx 0.11$ at $x^*\approx 0.71$, and
finally decreases as $\left(4 x^2\right)^{-1}$ for high frequencies
($x=\Omega\tau\gg 1$). Note that in order to use the single
miniband approximation and neglect the interminiband transitions, the
electric field frequency, in units of the 
energy $\hbar\Omega$, should be
less than the interminiband distance or, in other words, 
it should be of the order of or less than the miniband width
$\hbar\Omega\lesssim\Delta$.
The case $\hbar\Omega\simeq\Delta$ is the typical case in 
experiments \cite{SSL-exp} (see also our estimates above). Thus,
the dimensionless electric field, $\overline{E}$, (involved in 
the expression (\ref{result2})) is of the order of $e a E/\Delta$. 
To apply the Boltzmann equation to the description of miniband transport 
in a quantum superlattice,
the parameter $e a E/\Delta$ should always be small \cite{Bass1},
{\it i.e.}, in our notation, $\overline{E}\ll 1$. This remark shows
that
the DC generated in the SSL under the action of the high-frequency
electric field $E(t)$ is less than the corresponding current
$j=\sigma E$ generated by a constant bias of the same strength $E$ in the 
factor $\simeq 0.1 \overline{E}^2$. The numerical value of this factor is only
$\simeq 10^{-3}$ for typical experiments in SSLs, which implies that
the semiclassical approach is valid and at the same time gives serious
grounds for the observation of DC generation due to wave mixing.
\par
It is interesting  that in another scheme of 
rectification as suggested in Ref. \cite{Alekseev-prl}
only a single harmonic of a THz field is required.
However, in this case there is actually an {\it implicit} wave mixing. 
Indeed, even in the simplest case of rectification,
the self-consistent field generated in a SSL has a second harmonic of 
the fundamental frequency, $\Omega$,
so that the {\it total electric field} acting on the electrons 
has at least two
components with frequencies $\Omega$ and $2\Omega$ 
\cite{Alekseev-prl,Alekseev-prb}.
Consequently, the effect proposed in Ref. \cite{Alekseev-prl}
may arise due to the wave mixing of the induced self-consistent field
and the applied external THz field. This effect will be similar to 
the effect of DC
generation  arising at the wave mixing of two
commensurate THz harmonics as described in Ref.
\cite{Alekseev-prl,Goychuk-epl}.
 Thus, in spite of
obvious differences, the approaches to rectification of THz
radiation developed in Refs
\cite{Alekseev-prl,Goychuk-epl} are not contradictory but rather 
complement each other. 
\par
Now, let us consider the generation of DC at harmonic mixing 
in the III-V type semiconductors with narrow gaps, where the 
nonparobolicity is also strong.
Mixing of  mm-waves in n-doped InAs, InSb and GaAs
semiconductors has been studied experimentally in
\cite{Patel,Wynne,Gorky-exp} and discussed theoretically in
\cite{Wolff,Gorky-theor} a long time ago. However, we do not know 
of any reports devoted to the observation of DC generation at wave 
mixing in these semiconductors. This is probably because the experiments
\cite{Patel,Wynne,Gorky-exp} were
devoted only to the mixing of waves with similiar frequencies or third
harmonic generation. 
\par
Let us estimate the DC that could be generated in
such semiconductors at harmonic mixing. 
Semiconductors such as InAs, InSb can be described \cite{Wolff,Wynne}
by the Kane four-band model \cite{Kane,Bir}.
Of course, for a proper description of the electron-hole
dynamics we have to take into account all four bands,
which is, in fact, a very tedious task. 
However, in a weak field assuming that, due to dominant donor doping, 
the electron concentration is larger than the hole concentration,
we may take into account
only one of the energy bands: the electron branch with
the energy-momentum dispersion relation \cite{Kane,Bass2,Bir} 
in the form
%Eq(8)--------------------------
\begin{equation}
\label{kane}
\varepsilon(p)=\frac{\varepsilon_g}{2}\left[ \left(
1+\frac{2 p^2}{m\varepsilon_g}\right)^{1/2}-1\right],
\end{equation}
%------------------------------------
where $\varepsilon_g$ is the width of the gap.
Note that GaAs cannot be described by the dependence
(\ref{kane}), however in the weak field limit, the nonparabolicity, 
$\eta$, is roughly twice as great as follows from the Kane model
\cite{Wynne}. Note also that, due to a diamond structure 
of the III-V type semiconductors, into a nonparabolicity coefficient 
 the cubic invariants will also contribute, however,
these contributions should not exceed those which follow from 
the Kane model.
In the limiting case $p/\sqrt{m\varepsilon_g}\ll 1$, we have the
dependence (\ref{effective-mass}) with $\eta=-1/(m^2\varepsilon_g)$
and the DC generated at wave mixing (\ref{dc-ham}) is of the order of
%Eq(9)--------------------------
\begin{equation}
\label{dc-kane}
j_{\rm dc}\simeq\frac{3}{4} e n\left(\frac{\varepsilon_g}{m} \right)^{1/2}
{\tilde{E}}_1^2 {\tilde{E}}_2,\quad
{\tilde{E}}_l\equiv\frac{e E_l}{l\Omega\left(
m\varepsilon_g\right)^{1/2}},
\quad (l=1,2).
\end{equation}
%------------------------------------
The condition of weak field $p/(m\varepsilon_g)^{1/2}\ll 1$
takes the form ${\tilde{E}}_l\ll 1$ $(l=1,2)$.
For the donor doped n-InSb with the gap
 width $\varepsilon_g\approx 0.2$ eV, 
effective mass $m\approx 0.016 m_e$, level of doping $n\simeq
10^{16}$
${\rm cm^{-3}}$, sample area $S\simeq 10^{-4}$ ${\rm cm^2}$,
field frequency $\Omega\simeq 10^{11}$ ${\rm sec^{-1}}$ and
field strength of $\simeq 100$ V/cm, we get the DC
$I_{\rm dc}=j_{\rm dc} S$ of the order of several {\rm mA}. At the same
time, the condition $\tilde{E}\simeq 0.1\ll 1$ guarantees the
applicability of the weak field limit here.
\par
It is important to note  that
for gapless or narrow-gap semiconductors the current carriers
may be excited by
microwave radiation. That is their concentration may be changed
by the applied field.
On one hand the nonparabolicity of the energy spectrum
is larger in wide band semiconductors, while on the other hand
the concentration of electrons which may be excited by
microwave radiation may be larger in narrow gap or gapless
semiconductors or semimetals. Therefore, the most optimal conditions
for the described effect would be in semiconductors having narrow 
gaps and wide bands, like InSb and other analogous compounds.
\par
In this estimation of the DC in InSb
we actually neglect a scattering of the electrons by impurities, defects
and phonons;
which is in itself an important process in semiconductors. In this respect, we
should make two remarks. 
First of all, if the frequency $\Omega$ is of the same order as the
inverse characteristic relaxation time $\tau^{-1}$,
which is the case of experiments \cite{Patel,Wynne,Gorky-exp}, 
the DC should be of the same
order as estimated in Eq. (\ref{dc-kane}) \cite{comment}.
Otherwise, the dependence of the DC will have a form
similar to that obtained above for the case of a SSL, Eq. 
(\ref{result2}). 
\par
Secondly,
several other mechanisms of nonlinearity could also be responsible for 
wave mixing in semiconductors \cite{Patel,Wolff,Gorky-theor}. One of the
most important  of these is the so-called heating mechanism
where the  nonlinearity is
related to the dependence of a relaxation constant on the field
\cite{Gorky-exp,Gorky-theor}. This is mainly responsible for 
the mixing of mm-waves at low temperatures ($\simeq 4$K)
\cite{Gorky-exp}.
On other hand,
the experiments \cite{Patel,Wynne,Gorky-exp} and theories
\cite{Wolff,Gorky-theor} demonstrate that at temperatures $\simeq 80$K
the nonlinearity related to the nonparabolicity of the conduction band  becomes
more important. The estimates presented in \cite{Gorky-theor}
show that the relative influence of these two different types of nonlinearity is
dependent not only on the temperature but also on the type of wave
mixing.
For the mixing of waves with similiar frequencies (the case of experiments
\cite{Patel}), the nonparabolicity is the dominant nonlinearity for 
temperatures $\simeq 80$K. However, for the mixing of waves with an integer
ratio of frequencies, {\it e.g.} for a third harmonic generation,
the influence of both types of nonlinearity is comparable 
\cite{Gorky-exp,Gorky-theor}.
In any case, the effect of DC generation
arising due to the mixing of mm-waves should exist in narrow gap
III-V and even some II-VI semiconductors.  We hope
that our results describing this new effect
will attract the attention of experimentalists to this very intriguing
issue.
\par
In summary,  with the use of
 the semiclassical Boltzmann equation we have found the novel effect
of DC generation in semiconductors and
semiconductor microstructures driven
by two pure coherent electromagnetic waves of commensurate frequencies.
The described effect originates from the nonlinearity
associated with the nonparabolicity of the energy band and
therefore, it is universal and may be observed in any semiconductors.
Thus, our findings indicate that any semiconductor
may be considered as a particular type of nonlinear
medium serving as a generater of DC \cite{Breymayer}.
This effect is very different from the 
photogalvanic effect arising in semiconductors without a center of 
symmetry \cite{Belinicher}.
Finally, we  propose that experiments be set up for the 
mixing of millimetre- and submillimetre-waves in 
semiconductors and semiconductor superlattices
to observe the predicted generation of direct current. 
\par
We thank David Campbell, Ethan Cannon, Laurence Eaves,
Holger Grahn, Peter Kleinert, Eckehard Sch\"{o}ll, Andreas Wacker
for useful discussions on THz field driven semiconductor superlattices and
Yuri Romanov for bringing our attention to refs.
\cite{Romanov-opt,Romanov-ftt}.
KNA thanks the Department of Physics, Loughborough University for
hospitality
and Krasnoyarsk Regional Science Foundation for partial support.

\appendix
\section{}
\label{appendix1}
In this Appendix we will derive Eq. (\ref{j_dc-Rom}).
Starting from the Boltzmann equation for a distribution function 
$f(p,t)$
%Eq(A1)-----------------------------------------------------
\begin{equation}
\label{AA1}
\frac{\partial f}{\partial t}+e E \frac{\partial f}{\partial p}=
-\frac{f-f^{\rm eq}}{\tau}
\end{equation}
%------------------------------------------------------------
with an equilibrium distribution function $f^{\rm eq}(p)$ and
electric field $E(t)=E_1\cos(\omega_1 t)+E_2\cos(\omega_2 t+\phi)$,
Orlov and Romanov \cite{Romanov-ftt} have found the exact expression
for time-dependent current
%Eq(A2)-----------------------------------------------------
\begin{equation}
\label{AA2}
j(t)=e n \int v(p) f(p,t) {\rm d}p
\end{equation}
%-----------------------------------------------------------
in the form
%Eq(A3)-----------------------------------------------------
\begin{eqnarray}
\label{AA3}
\lefteqn{
j(t)=j_0 \Biggl[  \frac{i}{2}
\sum_{\mu_1,\mu_2=-\infty}^{+\infty}
\sum_{\nu_1,\nu_2=-\infty}^{+\infty}
J_{\mu_1}(\xi_1) J_{\mu_2}(\xi_2) J_{\mu_1+\nu_1}(\xi_1)
J_{\mu_2+\nu_2}(\xi_2)}\nonumber \\
 & &
\frac{1}{1+i(\mu_1\omega_1+\mu_2\omega_2)\tau}
 \exp\left( -i\nu_1\omega_2 t-i\nu_2 (\omega_2 t+\phi)\right) \Biggr]
+ {\rm c. c.},
\end{eqnarray}
%------------------------------------------------------------
where $\xi_k=\frac{e a E_k}{l\hbar\omega_k}$ ($k=1,2$) and
$j_0=\frac{\hbar\sigma}{e a\tau}$. 
In this appendix, we first present simple derivation\footnote{The
simplest way is to solve the Boltzmann equation in the Fourier
representation. For the case $E(t)=E_0\cos(\omega t)$, this method
is presented in clear form in: Bass F. G., Rubinshtein E. A.,
{\it Fiz. Tverd. Tela}, {\bf 19} (1977) 1379 [{\it Sov. Phys. Solid State},
{\bf 19} (1977) 800].
}
of Eq.(\ref{AA3}), and then will obtain the expression for rectified DC.
\par
Using periodicity of distribution function in $p-$space
%Eq(A4)-----------------------------------------------------
\begin{equation}
\label{AA4}
f(p,t)=\sum_{l=-\infty}^{+\infty} f_l(t)
\exp\left( i \frac{a p l}{\hbar}\right),
\end{equation}
%-----------------------------------------------------------
the Boltzmann equation for $l$th harmonics could be represented as
%Eq(A5)-----------------------------------------------------
\begin{equation}
\label{AA5}
\frac{\partial f_l}{\partial t}-\frac{i e l a}{\hbar c} 
\frac{\partial A}{\partial t} f_l=
-\frac{f_l-f_{l}^{\rm eq}}{\tau},\quad 
E=-\frac{1}{c}\frac{\partial A}{\partial t}.
\end{equation}
%-----------------------------------------------------------
Assuming $f(p,t=-\infty)=f^{\rm eq}(p)$, the solution of (\ref{AA5}) is
%Eq(A6)-----------------------------------------------------
\begin{equation}
\label{AA6}
f_l=\tau^{-1} \exp\left(-\frac{t}{\tau}\right)
\exp\left[\frac{i e l a A(t)}{\hbar c}\right] f_{l}^{\rm eq}
\int_{-\infty}^{t} \exp\left(\frac{t'}{\tau}\right)
\exp\left[-\frac{i e l a A(t')}{\hbar c}\right] {\rm d}t'.
\end{equation}
%-----------------------------------------------------------
This solution is valid for arbitrary $A(t)$; in our case
$A(t)=-\frac{E_1 c}{\omega_1}\sin(\omega_1 t)-\frac{E_2 c}{\omega_2}
\sin(\omega_2 t+\phi)$. Using the Bessel functions expansion
%Eq(A7)-----------------------------------------------------
\begin{equation}
\label{AA7}
\exp(\pm i\xi\sin\theta)=\sum_{m=-\infty}^{+\infty} J_m(\xi)
\exp(\pm i m\theta),
\end{equation}
%-----------------------------------------------------------
we can expand $\exp[\pm\frac{i e l a A(t)}{\hbar c}]$ in (\ref{AA6}) 
and get after integration:
%Eq(A8)-----------------------------------------------------
\begin{equation}
\label{AA8}
f_1=f_{1}^{\rm eq}\sum_{k_1,k_2}\sum_{n_1,n_2} 
J_{n_1}(\xi_1) J_{n_2}(\xi_2) J_{k_1}(\xi_1) J_{k_2}(\xi_2)
\frac{e^{i(n_1-k_1)\omega_1 t} e^{i(n_2-k_2)\omega_2 t} }
{1+i(n_1\omega_1+n_2\omega_2)\tau}
e^{i(n_2-k_2)\phi}.
\end{equation}
%-----------------------------------------------------------
Now, introducing new summation indexes $\nu_l=k_l-n_l$ ($l=1,2$)
and substituting Eq. (\ref{AA8}) into the expression for the current
$j=\frac{j_0}{2} i \left( f_1-f_{-1}\right)$,
we get the formula (\ref{AA3}). The Fourier transform of equilibrium
distribution function $f_{1}^{\rm eq}$ determines the temperature
dependence of conductivity.
\par
Finally, taking  $\omega_1=\Omega$, $\omega_2=2\Omega$ and averaging 
$j(t)$ [Eq. (\ref{AA3})]
over period of {\it ac} field $2\pi/\Omega$, we get for the rectified 
DC $j_{\rm dc}=\langle j(t)\rangle$ the formula (\ref{j_dc-Rom}).

\section{}
\label{appendix2}
In this Appendix we derive Eq. (\ref{result1}).
We start with the expression for DC $j_{dc}$, Eq. (\ref{j_dc-Rom}), 
which now is convenient to rewrite in the form
%Eq(B1)-----------------------------------------------------
\begin{equation}
\label{A1}
j_{dc}=j_0 \sum_{\mu_1,\mu_2=-\infty}^{+\infty}\sum_{\nu=-\infty}^{+\infty}
\left[ A(\mu_1,\mu_2,\nu) + B(\mu_1,\mu_2,\nu)\right]
J_{\mu_1}(\xi_1) J_{\mu_2}(\xi_2) J_{\mu_1-2\nu}(\xi_1) J_{\mu_2+\nu}(\xi_2),
\end{equation}
%------------------------------------------------------------
where
%Eq(B2)-----------------------------------------------------
\begin{equation}
\label{A2}
A(\mu_1,\mu_2,\nu)=\frac{(\mu_1+2\mu_2) x\cos(\nu\phi)}
{1+(\mu_1+2\mu_2)^2 x^2},\quad
B(\mu_1,\mu_2,\nu)=\frac{\sin(\nu\phi)}{1+(\mu_1+2\mu_2)^2 x^2}
\end{equation}
%-----------------------------------------------------------
Using the Bessel function property $J_{n}(\xi)\simeq\xi^n$ for $\xi\ll 1$,
we will consider only
nonvanishing lowest order in $\xi_{1,2}\ll 1$ terms in (\ref{A1}).
We start with terms of $[ \mu_1=0,\mu_2=0,\nu=\pm 1 ]$:
%Eq(B3)-----------------------------------------------------
\begin{eqnarray}
\label{A3}
& [ A(0,0,1)+B(0,0,1)] & J_0 (\xi_1) J_0 (\xi_2) J_{-2}(\xi_1) J_1(\xi_2) +
\nonumber \\
& [ A(0,0,-1)+B(0,0,-1)] & J_0(\xi_1) J_0(\xi_2) J_2(\xi_1) J_{-1}(\xi_2) =
\nonumber \\
& &  2\sin(\phi)  J_0(\xi_1) J_0(\xi_2) J_2(\xi_1) J_1(\xi_2).
\end{eqnarray}
%-----------------------------------------------------------
Term with $[ 0,-1,1]$ plus term with $[ 0,1,-1]$:
%Eq(B4)-----------------------------------------------------
\begin{eqnarray*}
& & [ A(0,-1,1)+B(0,-1,1)]  J_0(\xi_1) J_{-1}(\xi_2) J_{-2}(\xi_1)
J_0(\xi_2) + \\
& & [ A(0,1,-1)+B(0,1,-1)] J_0(\xi_1) J_1(\xi_2) J_2(\xi_1) J_0(\xi_2)=
\end{eqnarray*}
\begin{equation}
\label{A4}
2 \left[  \frac{2x\cos\phi}{1+4 x^2}-\frac{\sin\phi}{1+4 x^2} \right]
J_0(\xi_1) J_1(\xi_2) J_2(\xi_1) J_0(\xi_2).
\end{equation}
%-----------------------------------------------------------
Term with $[1,0,1]$ plus term with $[-1,0,-1]$:
%Eq(B5)-----------------------------------------------------
\begin{eqnarray*}
& & [ A(1,0,1)+B(1,0,1)] J_1(\xi_1) J_0(\xi_2) J_{-1}(\xi_1) J_1(\xi_2) +  \\
& & [ A(-1,0,-1)+B(-1,0,-1)] J_{-1}(\xi_1) J_0(\xi_2) J_1(\xi_1) J_{-1}
(\xi_2)=
\end{eqnarray*}
\begin{equation}
\label{A5}
-2\left[ \frac{x\cos\phi}{1+x^2}+\frac{\sin\phi}{1+x^2} \right]
J_1^2(\xi_1) J_0(\xi_2) J_1(\xi_2).
\end{equation}
%-----------------------------------------------------------
Term with $[1,-1,1]$ plus term with $[-1,1,-1]$:
%Eq(B6)-----------------------------------------------------
\begin{eqnarray*}
& & [ A(1,-1,1)+B(1,-1,1)] J_1(\xi_1) J_{-1}(\xi_2) J_{-1}(\xi_1) J_0(\xi_2) +
\\
& & [ A(-1,1,-1)+B(-1,1,-1)] J_{-1}(\xi_1) J_1(\xi_2) J_1(\xi_1) J_0(\xi_2)=
\end{eqnarray*}
\begin{equation}
\label{A6}
2\left[ \frac{-x\cos\phi}{1+x^2}+\frac{\sin\phi}{1+x^2} \right]
J_1^2(\xi_1) J_0(\xi_2) J_1(\xi_2).
\end{equation}
%-----------------------------------------------------------
Term with $[2,0,1]$ plus term with $[-2,0,-1]$:
%Eq(B7)-----------------------------------------------------
\begin{eqnarray*}
& & [ A(2,0,1)+B(2,0,1)] J_2(\xi_1) J_0(\xi_2) J_0(\xi_1) J_1(\xi_2) +\\
& & [ A(-2,0,-1)+B(-2,0,-1)] J_{-2}(\xi_1) J_0(\xi_2) J_0(\xi_1) J_{-1}
(\xi_2)=
\end{eqnarray*}
\begin{equation}
\label{A7}
2\left[ \frac{2x\cos\phi}{1+4x^2}+\frac{\sin\phi}{1+4x^2} \right]
J_2(\xi_1) J_0(\xi_2) J_0(\xi_1) J_1(\xi_2).
\end{equation}
%-----------------------------------------------------------
Term with $[2,-1,1]$ plus term with $[-2,1,-1]$:
%Eq(B8)-----------------------------------------------------
\begin{eqnarray*}
& & [ A(2,-1,1)+B(2,-1,1)] J_2(\xi_1) J_{-1}(\xi_2) J_0(\xi_1) J_1(\xi_2) +\\
& & [ A(-2,1,-1)+B(-2,1,-1)] J_{-2}(\xi_1) J_1(\xi_2) J_0(\xi_1) J_0(\xi_2)=
\end{eqnarray*}
\begin{equation}
\label{A8}
=-2\sin(\phi) J_2(\xi_1) J_1(\xi_2) J_0(\xi_1) J_0(\xi_2).
\end{equation}
%-----------------------------------------------------------
All other terms in Eq. (\ref{A1}) either compensate each other or
are of order which is greater than $\xi^3$.
Combining (\ref{A3})-(\ref{A8}), we get following formula for the DC
%Eq(B9)-----------------------------------------------------
$$
\frac{j_{dc}}{j_0}= \frac{8x\cos\phi}{1+4x^2} J_0(\xi_1) J_1(\xi_2) J_2(\xi_1)
J_0(\xi_2)-\frac{4x\cos\phi}{1+x^2} J_1^2(\xi_1) J_0(\xi_2) J_1(\xi_2)
$$
\begin{equation}
\label{A9}
\approx \frac{1}{2} \left(
\frac{x}{1+4x^2}-\frac{x}{1+x^2}\right) \xi_1^2 \xi_2 \cos\phi
=-\frac{3}{2}\frac{x^3}{4x^4+5x^2+1} \xi_1^2 \xi_2 \cos\phi,
\end{equation}
%-----------------------------------------------------------
where we have used the expansion
%Eq(B10)-----------------------------------------------------
\begin{equation}
\label{A10}
J_n(\xi)\approx (\xi/2)^n (1/n!),\quad \xi\ll 1.
\end{equation}
%-----------------------------------------------------------
As follows from Eq. (\ref{A9}), the absolute value of the DC as a
function of $x\equiv\Omega\tau$ at fixed $\xi_1$, $\xi_2$ and $\phi$
increases like $x^3$ for small $x$, then reaches maximum at $x^*\approx
1.3$ ($|j_{dc}(x=x^*)/j_0|\approx 0.16\xi_1^2 \xi_2 \cos\phi$), and
finally decreases as $3/(8 x)$  for large $x$.
\par
To check these calculations and understand what is the order in $\xi_{1,2}$
for the next terms in expression for DC, we perform the symbolic computations
for Eq. (\ref{A1}) with account of (\ref{A10}) using MAPLE.
As example, we present here the form of DC obtained by the summation in
(\ref{A1}) over intervals $-2\leq(\mu_1,\mu_2,\nu)\leq 2$:
%Eq(MAPLE)---------------------------------------------------
\begin{eqnarray*}
\lefteqn{\Biggl(\Biggl({\displaystyle \frac {1}{294912}} \,{\displaystyle
\frac {{\rm cos}(\phi )\,x}{1 + 4\,x^{2}}}  + {\displaystyle
\frac {1}{589824}} \,{\displaystyle \frac {{\rm sin}(\phi )}{1 +
4\,x^{2}}} \Biggr)\,\xi_2^{5}} \\
 & & \mbox{} + \Biggl({\displaystyle \frac {1}{4096}} \,{\displaystyle
\frac {{\rm cos}(\phi )\,x}{1 + 36\,x^{2}}}  - {\displaystyle
\frac {1}{24576}} \,{\displaystyle \frac {{\rm sin}(\phi )}{1 +
36\,x^{2}}}  + {\displaystyle \frac {1}{24576}} \,{\rm sin}(\phi
)\Biggr)\,\xi_2^{3} \\
 & & \mbox{} + \Biggl({\displaystyle \frac {1}{768}} \,{\displaystyle
\frac {{\rm cos}(\phi )\,x}{1 + 16\,x^{2}}}  - {\displaystyle
\frac {1}{3072}} \,{\displaystyle \frac {{\rm sin}(\phi )}{1 + 16
\,x^{2}}}  - {\displaystyle \frac {1}{1536}} \,{\displaystyle
\frac {{\rm cos}(\phi )\,x}{1 + 4\,x^{2}}}  + {\displaystyle
\frac {1}{3072}} \,{\displaystyle \frac {{\rm sin}(\phi )}{1 + 4
\,x^{2}}} \Biggr)\,\xi_2\Biggr)\xi_1^{6}\mbox{} + \Biggl( \\
 & & \Biggl({\displaystyle \frac {1}{6144}} \,{\displaystyle \frac {
{\rm cos}(\phi )\,x}{1 + 9\,x^{2}}}  + {\displaystyle \frac {1}{
18432}} \,{\displaystyle \frac {{\rm sin}(\phi )}{1 + 9\,x^{2}}}
\Biggr)\,\xi_2^{5} \\
 & & \mbox{} +\Biggl({\displaystyle \frac {1}{768}} \,{\displaystyle
\frac {{\rm cos}(\phi )\,x}{1 + x^{2}}}  + {\displaystyle \frac {
1}{768}} \,{\displaystyle \frac {{\rm sin}(\phi )}{1 + x^{2}}}
 + {\displaystyle \frac {5}{768}} \,{\displaystyle \frac {{\rm
cos}(\phi )\,x}{1 + 25\,x^{2}}}  - {\displaystyle \frac {1}{768}
} \,{\displaystyle \frac {{\rm sin}(\phi )}{1 + 25\,x^{2}}} \Biggr)\,
\xi_2^{3} \\
 & & \mbox{} +\Biggl({\displaystyle \frac {1}{32}} \,{\displaystyle
\frac {{\rm cos}(\phi )\,x}{1 + 9\,x^{2}}}  - {\displaystyle
\frac {1}{96}} \,{\displaystyle \frac {{\rm sin}(\phi )}{1 + 9\,x
^{2}}}  - {\displaystyle \frac {1}{96}} \,{\displaystyle \frac {
{\rm cos}(\phi )\,x}{1 + x^{2}}}  + {\displaystyle \frac {1}{96}
} \,{\displaystyle \frac {{\rm sin}(\phi )}{1 + x^{2}}}\Biggr)\,\xi_2\Biggr)
\xi_1^{4}\mbox{} + \Biggl(\Biggl({\displaystyle \frac {1}{256}} \,
{\displaystyle \frac {{\rm cos}(\phi )\,x}{1 + 36\,x^{2}}}  \\
 & & \mbox{} + {\displaystyle \frac {1}{1536}} \,{\displaystyle
\frac {{\rm sin}(\phi )}{1 + 36\,x^{2}}}  - {\displaystyle
\frac {5}{768}} \,{\displaystyle \frac {{\rm cos}(\phi )\,x}{1 +
25\,x^{2}}}  - {\displaystyle \frac {1}{768}} \,{\displaystyle
\frac {{\rm sin}(\phi )}{1 + 25\,x^{2}}}  + {\displaystyle
\frac {1}{384}} \,{\displaystyle \frac {{\rm cos}(\phi )\,x}{1 +
16\,x^{2}}}  + {\displaystyle \frac {1}{1536}} \,{\displaystyle
\frac {{\rm sin}(\phi )}{1 + 16\,x^{2}}} \Biggr) \\
 & & \xi_2^{5}\mbox{} + \Biggl( - {\displaystyle \frac {3}{16}} \,
{\displaystyle \frac {{\rm cos}(\phi )\,x}{1 + 9\,x^{2}}}  +
{\displaystyle \frac {1}{8}} \,{\displaystyle \frac {{\rm cos}(
\phi )\,x}{1 + 16\,x^{2}}}  + {\displaystyle \frac {1}{16}} \,
{\displaystyle \frac {{\rm cos}(\phi )\,x}{1 + 4\,x^{2}}}\Biggr)\,\xi_2^{3}
 +\Biggl({\displaystyle \frac {1}{2}} \,{\displaystyle \frac {
{\rm cos}(\phi )\,x}{1 + 4\,x^{2}}}  - {\displaystyle \frac {1}{2
}} \,{\displaystyle \frac {{\rm cos}(\phi )\,x}{1 + x^{2}}} \Biggr)\,
\xi_2\Biggr)\xi_1^{2}
\end{eqnarray*}
%----------------------------------------------------------------
As is evident from this result, next after (\ref{A9}) contribution
do DC is of order of $\xi^5$ ({\it i.e.}, of order of $\xi_1^l\xi_2^k$
with $k+l=5$).

\begin{figure}
\caption{
The dependence of DC on
amplitudes  $E_1$ and $E_2$ [Eq. (5)].
%[Eq. (\ref{j_dc-Rom})].
The values $E_1$ and $E_2$ are expressed in units of
$\hbar \omega/(e a)$ and $2 \hbar\omega/(ea)$, respectively,
and DC is normalized to $j_0$.
The product $\Omega\tau$ has the
values: $x = 0.01$ (a), $x=0.2$ (b), and $x=1.0$ (c). Phase $\phi=0$.}
\end{figure}

\begin{references}

\bibitem[a]{email1}
E-mail: kna@vist.krascience.rssi.ru, kna@iph.krasnoyarsk.su

\bibitem[b]{email2}
E-mail: F.Kusmartsev@lboro.ac.uk

%1
\bibitem{Esaki-ibm}
Esaki L. and Tsu R., {\it IBM J. Res. Dev.}, {\bf 14} (1970) 61.

%2
\bibitem{THz-emission}
Bouchard A. M. and Luban M., {\it Phys. Rev. B}, {\bf 47} (1993) 6815;
Martini R. {\it et al.}, {\it Phys. Rev. B}, {\bf 54} (1996) R14325.

%3
\bibitem{Bass1}
Bass F. G. and Tetervov A. P., {\it Phys. Rep.}, {\bf 140} (1986) 237.

%4
\bibitem{Bass2}
Bass F. G. and Vatova L. B., {\it Phys. Rep.}, {\bf 241} (1994) 219.

%5
\bibitem{SSL-exp}
Keay B. J. {\it et al.}, {\it Phys. Rev. Lett.}, {\bf 75} (1995) 4098;
{\bf 75}, (1995) 4102; Unterrainer K. {\it et al.},
{\it Phys. Rev. Lett.}, {\bf 76} (1996) 2973; Schomburg E. {\it et al.},
{\it Appl. Phys. Lett.}, {\bf 68} (1996) 1096; Zeuner S. {\it et al.},
{\it Phys. Rev. B}, {\bf 53} (1996) R1717; Ignatov A. A. {\it et al.},
{\it Ann. der Physik (Leipzig}, {\bf 3} (1994) 137; Winnerl S. {\it et
al.}, {\it Phys. Rev. B}, {\bf 56} (1997) 10303.

%6
\bibitem{Esaki-apl}
Esaki L. and Tsu R., {\it Appl. Phys. Lett.}, {\bf 19} (1971) 246.

%7
\bibitem{Romanov-opt}
Romanov Yu. A., {\it Optika i Spektr.}, {\bf 33} (1972) 917
({\it Sov. Phys. Optics and Spectroscopy}).

%8
\bibitem{Romanov-ftt}
Orlov L. K. and Romanov Yu. A., {\it Fiz. Tverd. Tela} {\bf 19} (1977)
726 ({\it Sov Phys. Solid State});
Romanov Yu. A., Orlov L. K. and Bovin V. P., {\it Fiz. Tekhn. Polupr.}
{\bf 9}, (1978) 1665 ({\it Sov. Phys. Semicond.}).

%9
\bibitem{Alekseev-prl}
Alekseev K. N. {\it et al.}
{\it Phys. Rev. Lett.}, {\bf 80} (1998) 2669 [ also cond-mat/9709026 ]

%10
\bibitem{Alekseev-prb}
Alekseev K. N. {\it et al.}, {\it Phys. Rev. B}, {\bf 54} (1996) 10625
[ also cond-mat/9604173 ];
{\it Physica D}, {\bf 113} (1998) 129.

%11
\bibitem{Goychuk-epl}
Goychuk I. and H\"{a}nggi P., {\it Europhys. Lett.}, {\bf 43} (1998)
503.

%12
\bibitem{Goychuk-prl}
Goychuk I., Grifoni M. and H\"{a}nggi P.,
{\it Phys. Rev. Lett.}, {\bf 81} (1998) 649.

%13
\bibitem{Bir} 
Bir G.L. and Pikus G. E., 
{\it Symmetry and deformational effects in semiconductors}, (Nauka, 
Moscow) 1972.

%14
\bibitem{Patel}
Patel C. K. N., Slusher R. E. and Fleury P. A.,
{\it Phys. Rev. Lett.}, {\bf 17} (1966) 1011.

%15
\bibitem{Wynne}
Wynne J. J., {\it Phys. Rev.}, {\bf 178} (1969) 1295.

%16
\bibitem{Gorky-exp}
Belyantsev A. M. {\it et al.},
%%, Valov V. A., Genkin V. N., Leonov A. M. and Trifonov B. A.,
{\it Zh. Eksp. Teor. Fiz.}, {\bf 61} (1971) 886
({\it Sov. Phys. JETP});
Genkin V. N., Kozlov V. A. and Piskarev  V. I.,
{\it Fiz. Tekh. Polupr.}, {\bf 8} (1974) 2013
({\it Sov. Phys. Semicond.}).

%17
\bibitem{Wolff}
Wolff P. A. and Pearson G. A., {\it Phys. Rev. Lett.}, {\bf 17} (1966)
1015.

%18
\bibitem{Gorky-theor}
Belyantsev A. M., Kozlov  V. A. and Trifonov B. A.,
{\it phys. status sol. (b)}, {\bf 48} (1971) 581.

%19
\bibitem{Kane}
Kane E. O., {\it J. Phys. Chem. Solid.}, {\bf 1} (1957) 249.

%20
\bibitem{comment}
Alekseev K. N. and Kusmartsev F. V., (unpublished).

%21
\bibitem{Grahn-prb}
Grahn H. T., von Klitzing K., Ploog K. and D\"{o}hler G. H.,
{\it Phys. Rev. B}, {\bf 43} (1991) 12094; Sibille A. {\it et al.},
{\it Superlatt. Microstr.}, {\bf 13} (1993) 247.

%22
\bibitem{formula}
See also ref. \cite{Bass1}, p. 310, Eq (8.15).


%
\bibitem{Breymayer}
Breymayer H.-J., Risken H., Vollmer H. D. and Wonneberger W.,
{\it Appl. Phys. B}, {\bf 28} (1982) 335.

%
\bibitem{Belinicher}
Belinicher V. I. and Sturman B. I.,
{\it Uspekhi. Fiz. Nauk}, {\bf 130} (1980) 415
({\it Soviet Phys. Uspekhi}, {\bf 23} (1980) 199).

\end{references}
\end{document}